\documentclass[a4paper,12pt]{article}
\usepackage{graphicx} % Required for inserting images
\usepackage[margin=1in]{geometry} 
\usepackage{amsmath}

\usepackage{authblk} % for affiliations
\usepackage{setspace} % optional, for spacing

\title{\textbf{Multimode Photon-Photon Coupling}}

\author[1]{\small Shourya Viren\thanks{These authors contributed equally to this work.}}
\author[2]{Rakesh Kumar Nayak\textsuperscript{*}}
\author[2]{Biswanath Bhoi}
\author[2]{Rajeev Singh\thanks{Corresponding author: \texttt{rajeevs.phy@iitbhu.ac.in}}}

\affil[1]{Global City International School, Bengaluru, Karnataka-560075, India}
\affil[2]{Department of Physics, Indian Institute of Technology (Banaras Hindu University)\\Varanasi - 221005, India}
\date{}
\begin{document}

\maketitle
\section*{\centering Abstract}
This study investigates a planar hybrid system consisting of three complementary split-ring resonators (CSRRs), designed to examine interactions among multiple photon modes at room temperature. The system was modeled and simulated using the full-wave electromagnetic solver CST Microwave Studio. Analysis of the transmission spectra (\(|S_{21}|\)) as a function of frequency for different CSRR dimensions revealed distinct anti-crossing behavior, indicative of strong photon-photon coupling (PPC). To explain this phenomenon, we present theoretical framework that quantitatively captures the observed mode hybridization and provides estimates of the coupling strength (\(\Delta\)), which are further validated experimentally. This work not only elucidates the fundamental dynamics of PPC in planar systems but also offers practical guidance for designing hybrid platforms with tunable photon interactions, paving the way for future advancements in planar magnonic and hybrid photonic technologies.

\section{Introduction}
The advancement of quantum technologies has established quantum sensing, quantum processing, and quantum information as transformative paradigms that fundamentally redefine the boundaries of precision measurement, computational capability, and information transfer\cite{Degen2017}. Central to these developments is the precise manipulation and control of electromagnetic wave phenomena, which constitutes the theoretical and experimental foundation for achieving coherent quantum states and maintaining quantum coherence across diverse physical platforms. Quantum sensing exploits quantum mechanical effects such as superposition and entanglement to surpass the standard quantum limit, enabling ultrasensitive detection of physical parameters including magnetic field fluctuations, phase variations, and molecular dynamics with unprecedented precision and spatial resolution.\cite{Lezama1998,Bhoi2022,li2018} Concurrently, quantum processing harnesses quantum interference and quantum parallelism to execute algorithmic operations that exhibit exponential advantages over classical computational approaches, while quantum information protocols utilize quantum correlations and non-local quantum states to enable secure cryptographic protocols, distributed quantum computing architectures, and quantum communication networks. The convergence of these quantum domains has stimulated extensive theoretical and experimental investigations into hybrid quantum systems, wherein coherent coupling mechanisms between disparate physical subsystems—encompassing superconducting resonators, atomic ensembles, photonic structures, and spin systems—facilitate the development of integrated quantum platforms with enhanced operational capabilities and deterministic quantum state control. Despite the substantial progress in quantum technologies, significant challenges persist in developing scalable, planar quantum devices that can simultaneously exhibit both coupling-induced transparency (CIT) and coupling-induced absorption (CIA) phenomena with precise parametric control.\cite{Bhoi2022} While conventional approaches utilizing magnon-photon coupling systems have demonstrated remarkable achievements in quantum coherence manipulation, their implementation often requires complex three-dimensional architectures, cryogenic operating conditions, and sophisticated fabrication processes that limit practical scalability and integration into on-chip quantum systems\cite{Rao2021}. Furthermore, existing photon-photon coupling (PPC) configurations predominantly focus on single-mode operations or lack the systematic tunability necessary for dynamic control over coupling strengths and resonant behaviours. The absence of comprehensive theoretical frameworks that can accurately predict and control the transition between CIT and CIA regimes in planar metamaterial structures represents a critical gap that impedes the optimization of quantum device performance and the realization of multifunctional quantum photonic circuits. In this context, complementary split-ring resonator (CSRR) integrated with microstrip transmission lines emerges as a promising paradigm for addressing these fundamental limitations through its inherent advantages in quantum sensing applications, compact device integration, and parametric controllability\cite{bernier2018,harder2017}. CSRR-based metamaterial structures offer exceptional miniaturization capabilities for various microwave devices including filters, sensors, and photonic components, while microstrip and coplanar waveguide resonators have established their prominence in quantum optics and quantum information processing experiments, particularly in circuit QED applications\cite{metelmann2014,scully1997,tiwari2023}. The planar geometry of CSRR-microstrip hybrid systems facilitates straightforward fabrication processes, enables operation across a broad frequency spectrum from GHz to THz regimes, and provides systematic geometric tunability through structural parameter modifications\cite{liu2019,li2018,li2022}. Moreover, the emerging field of quantum metamaterials has demonstrated significant potential for quantum information science applications, positioning CSRR-based photon-photon coupling systems as viable candidates for next-generation quantum technologies that require both transparency and absorption control within a single, integrated platform.\cite{Nayak2025,Bhoi2022,Shrivastava2024} This work presents a comprehensive investigation of CSRR-microstrip photon-photon three mode coupling, demonstrating systematic control over coupling strengths through change in dimensions of the resonator system and establishing a classical theoretical framework that accurately predicts the observed CIT and CIA phenomena, thereby bridging the gap between fundamental quantum coupling mechanisms and practical quantum device implementations\cite{Degen2017,castel2017}.

\section{Design and Simulation Methodology }
The electromagnetic characteristics of the complementary split ring resonator (CSRR) coupled with a microstrip transmission line were systematically investigated using CST Microwave Studio, a full-wave electromagnetic solver. This configuration enables comprehensive analysis of photon mode excitations and photon-photon coupling phenomena between these complementary structures. The proposed design consists of a CSRR system precisely etched onto a 40 mm × 40 mm copper ground plane with a microstrip line strategically positioned to facilitate electromagnetic interactions. The ground plane is fabricated from high-conductivity copper with a uniform thickness of 0.035 mm. A dielectric substrate with thickness 0.8 mm and relative permittivity $\varepsilon_{r} = 4.4$ separates the microstrip line from the ground plane, providing the necessary electromagnetic field confinement and coupling. The CSRR structure is precisely oriented on the ground plane such that its gaps (each measuring 1 mm) are symmetrically aligned along the X-axis, which is orthogonal to the Y-axis where the microstrip line runs. To optimize coupling efficiency, the microstrip line is positioned with a vertical offset t = 1 mm from the central axis of the CSRR. The microstrip line is designed with a width of 1.45 mm and a thickness of 0.035 mm to achieve the standard 50 characteristic impedance, ensuring compatibility with conventional RF measurement systems\cite{zhang2016,zheng2023} The resonant properties of the system were parametrically investigated by varying the outer width L of the CSRR-B and CSRR-C structures across a range of 4 mm to 18 mm, while maintaining the outer width m of the CSRR-A constant at 8 mm. Vector Network Analyzer (VNA) ports were implemented at both ends of the microstrip line, as illustrated in Fig.1, enabling measurement of S-parameters across the frequency spectrum. This measurement configuration allows for thorough characterization of transmission and reflection coefficients, providing quantitative insights into the mode splitting, avoided crossings, and electromagnetically induced transparency/absorption analog behaviors of the CSRR-microstrip system\cite{walls1997}. The simulation methodology incorporates accurate modelling of material properties, mesh optimization, and appropriate boundary conditions to ensure computational efficiency while maintaining high fidelity in resolving the electromagnetic field distributions at the CSRR-microstrip interface.
\begin{figure}[h!]
    \centering
    \includegraphics[width=0.6 \textwidth]{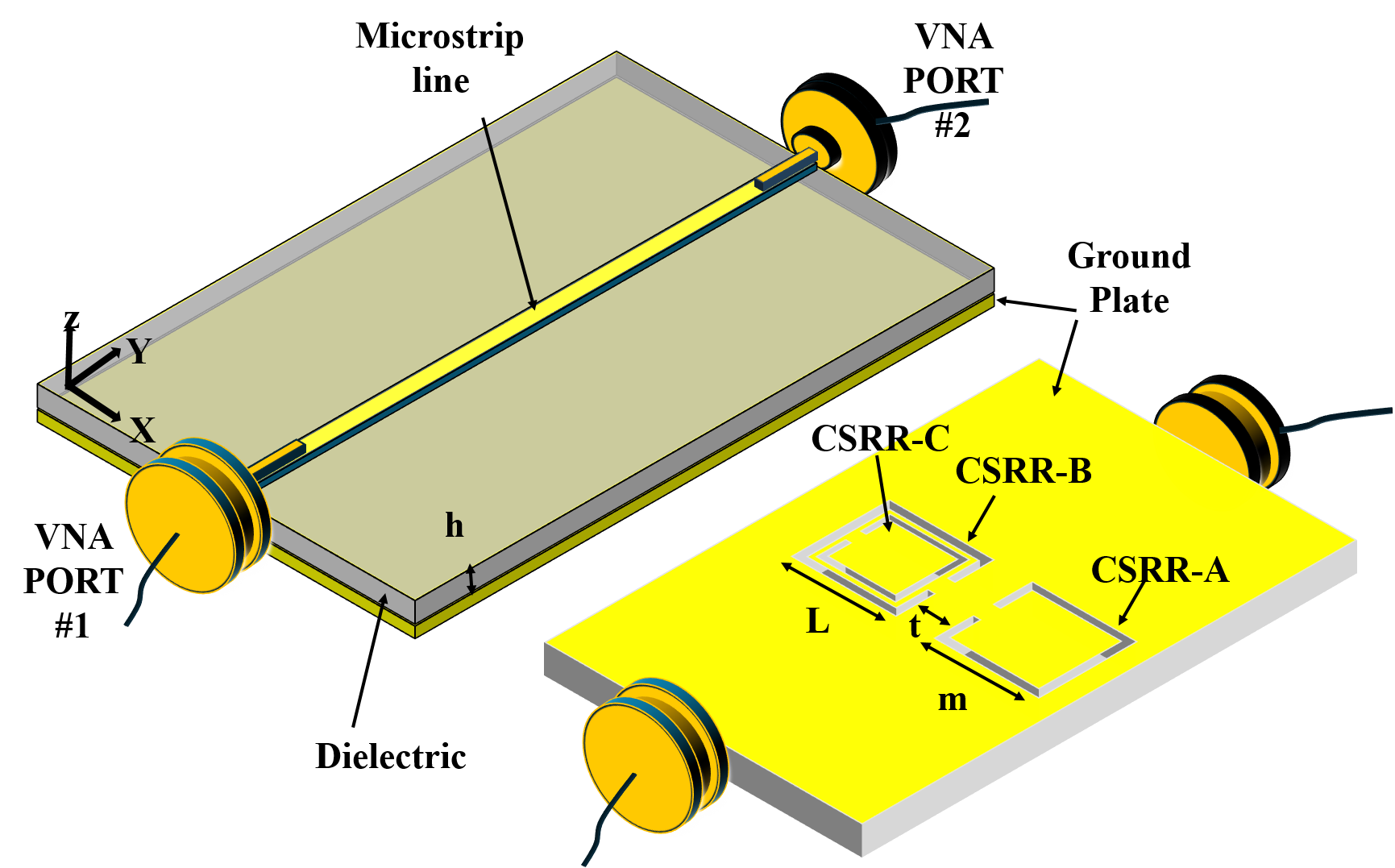}
    \caption{The simulation framework designed to study photon mode excitation is based on a planar geometry featuring three Complementary Split-Ring Resonator (CSRR) structures integrated with a microstrip feed line. Ports 1 and 2 of the feed line are connected to a Vector Network Analyzer (VNA) for characterization. The entire structure is fabricated on a dielectric substrate. A schematic illustration of the simulation setup for photon-photon coupling (PPC) is presented, showing both the front and back sides of the design.}
    \label{fig1}
\end{figure}

\section{Result and discussion }
To investigate the interaction mechanisms between CSRR and microstrip line resonators, we initially examined the transmission spectra $|S_{21}|$ as a function of frequency for single resonator systems, specifically analysing the CSRR-B and CSRR-C configurations coupled with the microstrip line. This analysis shows that the frequency of CSRR- B and CSRR- C vary with varying the side L, as shown in Fig.2. This systematic approach enabled comprehensive characterization of the individual electromagnetic behaviours of both concentric resonator elements, as well as their collective behaviour when integrated into the hybrid three-mode configuration.
\begin{figure}[h!]
    \centering
    \includegraphics[width=0.9 \textwidth]{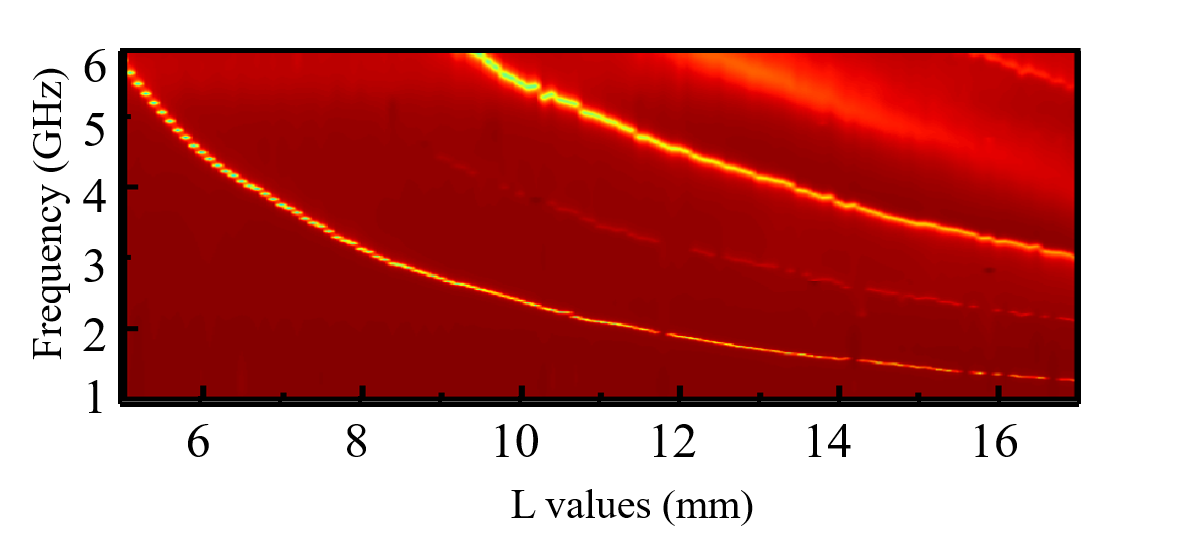}
    \caption{The transmission spectra of the CSRR-B and CSRR-C with microstrip line are illustrated using a colormap, highlighting the frequency-dependent response of the structure. The left and right frequency response curves correspond to CSRR-B and CSRR-C, respectively.}
    \label{fig2}
\end{figure}
\subsection{Observation of Anti-Crossing Phenomena }
Fig.3 presents the stacked transmission spectra $|S_{21}|$ of the CSRR-microstrip hybrid system as a function of frequency for varying CSRR dimensions L = 4mm to 18 mm, with gap = 0.1 mm and m = 8, revealing the intricate three-mode photon-photon coupling dynamics. The stacked plot representation provides a comprehensive visualization of the frequency evolution across multiple resonance modes, enabling clear identification of coupling-induced phenomena and mode interactions across the entire parameter space\cite{Bhoi2022,bhoi2014}. This representation facilitates the systematic tracking of resonance peaks as they undergo frequency shifts, amplitude modulations, and complex hybridization processes that are characteristic signatures of strong photon-photon coupling in multi-resonator systems. Three distinct resonance peaks are observed, each exhibiting unique frequency-dependent behaviours that provide direct evidence of coherent and dissipative  photon-photon coupling between the CSRR modes and the microstrip line resonance. 
\begin{figure}[h!]
    \centering
    \includegraphics[width=0.5 \textwidth]{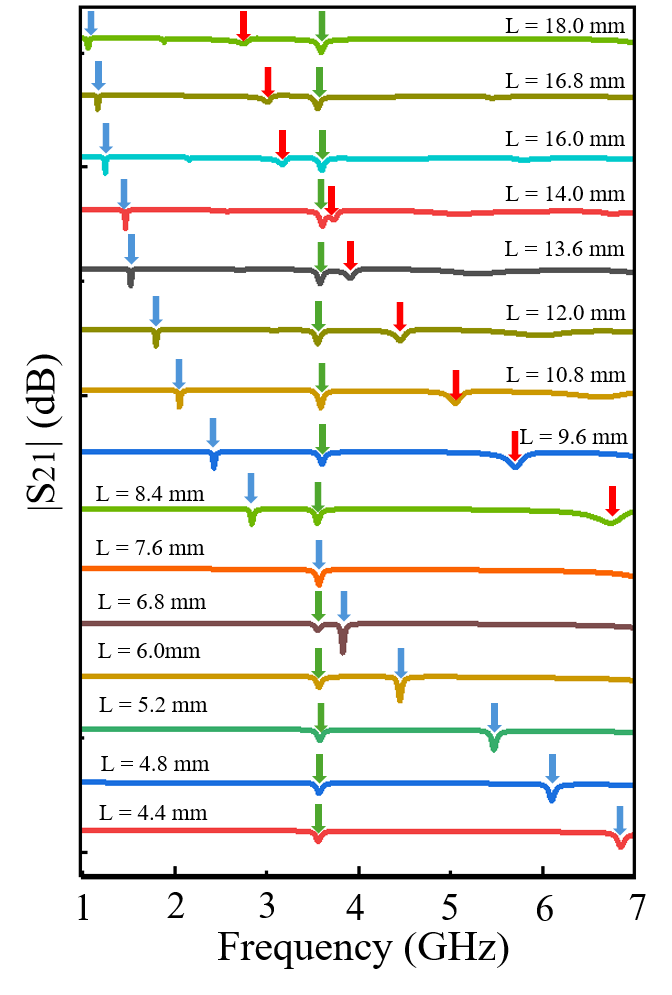}
    \caption{The stacked plot of transmission spectrum  $|S_{21}|$ plotted as a function of microwave frequency for varying CSRR sizes (L), providing a clear visualization of the dynamic response of the planar hybrid system. The three constituent resonators—CSRR-A, CSRR-B, and CSRR-C responses shown with marked green, blue, and red markers, respectively. This representation illustrates the interaction mechanisms among the three photon modes, including frequency shifts, amplitude modulations, and hybridization effects that are characteristic of strong photon–photon coupling in multi-resonator systems.}
    \label{fig3}
\end{figure}
The resonance peak indicated by green arrows corresponds to CSRR mode A, since the dimension of CSRR-A is kept fixed so its resonance behavior remain unchanged throughout the interaction. The resonance peaks marked by red and blue arrows represent CSRR modes B and C respectively. 

The red arrows (CSRR- B) indicate the outer ring resonance, given by $\omega_{B} = 1/\sqrt{L_{B}C_{B}}$ which is operating at relatively lower frequencies due to larger physical dimensions and correspondingly higher inductance characteristics. Conversely, the blue arrows (CSRR C) correspond to the inner ring resonance operating at higher frequencies because of reduced physical dimensions and lower effective inductance. Notably,  as the resonances of CSRR-B or CSRR-C approach that of CSRR-A, the amplitudes of the CSRR-B and CSRR-C peaks increase while the amplitude of the CSRR-A peak decreases. Once the resonances move apart, their original amplitudes are recovered. This reciprocal modulation of peak intensities reflects energy exchange between the coupled resonators. Consequently, two distinct coupling regions are observed, each giving rise to hybrid modes that represent superpositions of the original CSRR-A with CSRR-B and CSRR-C resonances, as clearly depicted in the colormap of Fig.4. In the region of L= 6 mm to 9 mm, the energy exchange  between CSRR-A and CSRR-B is coherent giving level repulsion and in the region of L=13 mm to 16 mm there is dissipative coupling giving level attraction. The most significant observation in the stacked plot is the clear manifestation of anti-crossing phenomena, where the resonance modes exhibit level repulsion behavior as they approach degeneracy conditions. As the system dimensions vary, these three distinct photon modes approach resonant coupling conditions at specific parameter values, resulting in the formation of complex multi-mode interactions characterized by avoided crossings rather than direct intersections. This anti-crossing behavior serves as unambiguous evidence of strong coherent coupling between the photon modes, indicating energy level hybridization and the formation of new collective eigenmodes of the coupled system.
\begin{figure}[h!]
    \centering
    \includegraphics[width=0.8 \textwidth]{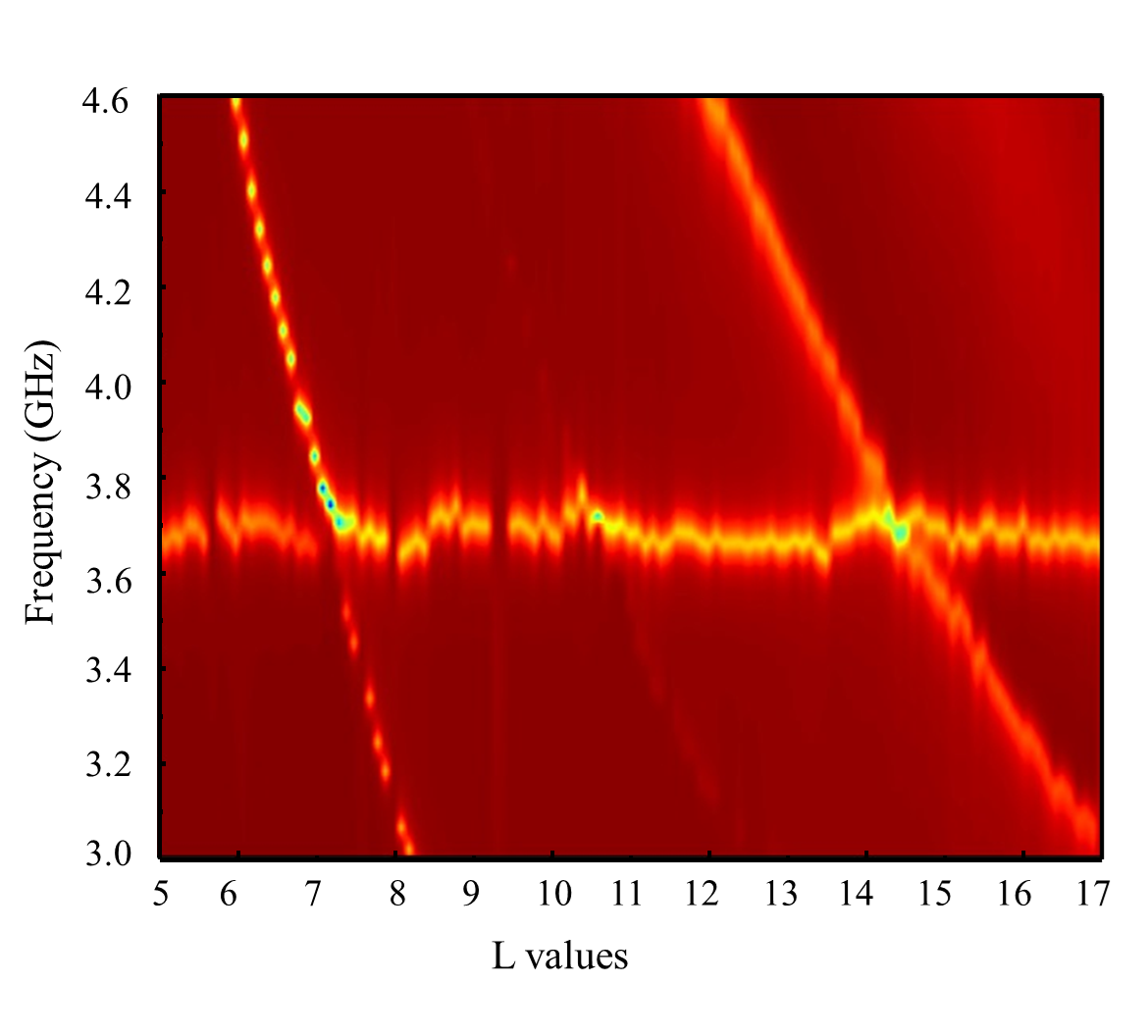}
    \caption{The transmission spectra of the three-mode CSRR-A, CSRR-B, and CSRR-C hybrid system are illustrated using a colormap, which highlights the frequency-dependent response of the structure. The spectra reveal two anticrossing regions: one in the range of L= 6mm to 9 mm, exhibiting level repulsion, and another in the range of L=13mm to 16 mm, exhibiting level attraction.}
     \label{fig4}
\end{figure}
The CSRR A mode serves as the primary coupling mediator, facilitating energy exchange between the concentrically arranged outer ring (CSRR B) and inner ring (CSRR C) modes through electromagnetic field overlap and mutual inductance effects. The amplitude modulation observed across different modes as they approach resonant coupling conditions demonstrates coherent energy redistribution, where the  mode CSRR-A exhibits the most dramatic amplitude variations, while the concentric modes (CSRR B and C) show correlated but less pronounced intensity changes.
\subsection{Theoretical model and fitting}
This study presents a theoretical framework based on Lagrangian mechanics and coupled-mode theory to describe the interactions within a three-mode CSRR-microstrip system. We derive the equations of motion governing the system, which incorporates complex eigenfrequencies and coupling constants for a more comprehensive representation of the resonator interactions. We begin by defining the system Lagrangian, incorporating kinetic and potential energy contributions:\[
T = \frac{1}{2} L_{A} \dot{q}_{A}^{2} 
  + \frac{1}{2} L_{B} \dot{q}_{B}^{2} 
  + \frac{1}{2} L_{C} \dot{q}_{C}^{2} 
  + M_{AB} \dot{q}_{A}\dot{q}_{B} 
  + M_{BC} \dot{q}_{B}\dot{q}_{C} 
  + M_{CA} \dot{q}_{C}\dot{q}_{A} 
  \tag {1}
\]
\[
U = \frac{1}{2} \left( \frac{q_A^2}{C_A} + \frac{q_B^2}{C_B} + \frac{q_C^2}{C_C} \right) 
\tag{2}
\]
where, $L_A, L_B, L_C$ and $C_A, C_B, C_C$ are the associated inductances and capacitances
of resonators (CSRR) A, B, and C. $M_{AB}, M_{BC}, M_{CA}$ are the mutual inductances among them. 
$q_i$ are generalized coordinates defined as 
\[
\dot{q}_i = \frac{dq_i}{dt}, \quad i \in \{A,B,C\}.
\]
Thus, the Lagrangian of the system is given by
\[
L = \tfrac{1}{2} \left( L_A \dot{q}_A^2 - \frac{q_A^2}{C_A} 
+ L_B \dot{q}_B^2 - \frac{q_B^2}{C_B} 
+ L_C \dot{q}_C^2 - \frac{q_C^2}{C_C} \right)
+ M_{AB} \dot{q}_A \dot{q}_B
+ M_{BC} \dot{q}_B \dot{q}_C
+ M_{CA} \dot{q}_C \dot{q}_A
\tag{3}
\]
From the Euler–Lagrange equation
\[
\frac{d}{dt}\left(\frac{\partial L}{\partial \dot{q}_i}\right) 
- \frac{\partial L}{\partial q_i} = 0
\tag{4}
\]
we obtain the equations of motion:
\[
L_A \ddot{q}_A + M_{AB}\ddot{q}_B + M_{CA}\ddot{q}_C + k_A q_A = 0
\tag{5}
\]
\[
L_B \ddot{q}_B + M_{AB}\ddot{q}_A + M_{BC}\ddot{q}_C + k_B q_B = 0
\tag{6}
\]
\[
L_C \ddot{q}_C + M_{AC}\ddot{q}_A + M_{BC}\ddot{q}_B + k_C q_C = 0
\tag{7}
\]
where $k_i = \tfrac{1}{C_i}$.  
Using the general solution $q_i(t) = Q_i e^{i\omega t}$ and rearranging terms, the system 
can be expressed in matrix form as $MQ = 0$, where $M$ and $Q$ are defined as:

\[
M = \begin{bmatrix}
k_A - L_A \omega^2 & -M_{AB}\omega^2 & -M_{AC}\omega^2 \\
-M_{AB}\omega^2 & k_B - L_B \omega^2 & -M_{BC}\omega^2 \\
-M_{AC}\omega^2 & -M_{BC}\omega^2 & k_C - L_C \omega^2
\end{bmatrix}
\tag{8}
\]

\[
Q = \begin{bmatrix}
Q_A \\ Q_B \\ Q_C
\end{bmatrix}
\tag{9}
\]
After substituting $\omega_i^2 = \tfrac{1}{L_i C_i}$ and replacing all diagonal terms 
with complex frequencies $\tilde{\omega}_A = \omega_A - i(\alpha+\gamma)$, 
$\tilde{\omega}_B = \omega_B - i(\beta+\gamma)$, 
$\tilde{\omega}_C = \omega_C - i(\kappa+\gamma)$,  
the above equation simplifies to yield the eigenvalue equations for the two coupling regions.
\begin{figure}[h!]
    \centering
    \includegraphics[width=0.8 \textwidth]{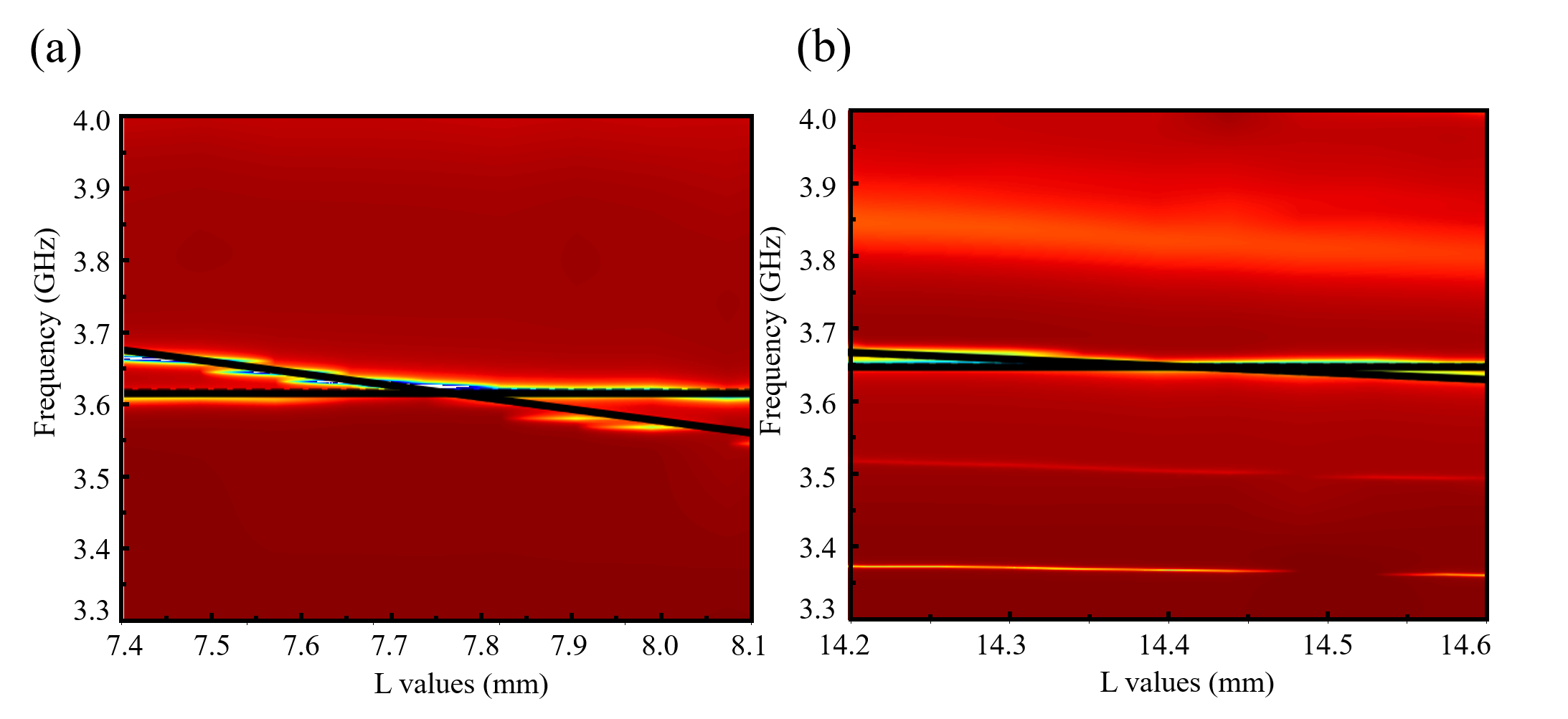}
    \caption{The theoretical fitting (black solid line) overlaid on the simulation data is shown for (a) the region L= 7.4 mm to 8.1 mm and (b) the region L= 14.2 to 14.6 mm.}
    \label{fig5}
\end{figure}
To quantitatively validate the coupling dynamics observed in the CST simulations, the exported $|S_{21}|$ data were fitted using the eigenvalue equation to analyze the interactions of CSRR-A with CSRR-B and CSRR-C, shown in Fig.5. The model incorporated the intrinsic resonance frequencies, with intrinsic dampings held fixed at $\kappa = 0.03290$, $\alpha = 0.02387$, and $\beta = 0.03579$, together with the extrinsic damping $\gamma$ and the inter-resonator couplings $\Delta_{AB}$, $\Delta_{BC}$, and $\Delta_{CA}$ confirming that all three oscillators participate with comparable strength and validating the three-mode interaction picture. The CIT regime i.e. level repulsion, exhibits coherent energy exchange between modes while in the CIA regime i.e. level attraction, dissipation dominates the system dynamics, which leads to merged resonances accompanied by a pronounced absorption dip, confirming close agreement between the model and simulation. Overall, these fits clearly demonstrate the coherent energy exchange (CIT) and dissipative coupling (CIA) and provide a concise set of parameters that define the tunability of the CSRR–microstrip photonic platform.

\section{Conclusion}
In conclusion, we have demonstrated three-mode photon–photon coupling in a planar CSRR–microstrip hybrid system, achieving electromagnetic transparency and absorption through systematic optimization of geometric parameters. Our experimental results confirm strong coupling between CSRR-A and the concentrically arranged CSRR-B and CSRR-C modes, as evidenced by clear anticrossing behavior with both level repulsion and level attraction observed in the transmission spectra. Two distinct operational regimes are identified: one in the range of L= 6mm to 9 mm, exhibiting level repulsion, and another in the range of L=13mm to 16 mm, exhibiting level attraction. This establishes the dual-functional nature of the metamaterial platform, enabling dynamic switching between transparency and absorption within a single integrated structure. The theoretical framework, based on Lagrangian mechanics and expressed through a characteristic matrix with complex eigenfrequencies and normalized coupling constants, accurately reproduces the experimental results and provides fundamental insights into the three-mode coupling dynamics. Furthermore, the planar architecture offers practical advantages such as simplified fabrication, room-temperature operation, and broad frequency tunability from the GHz to THz regimes, positioning CSRR-based metamaterials as strong candidates for next-generation quantum sensing, reconfigurable photonic circuits, and quantum information processing systems.
\section*{Acknowledgment}
The work was supported by the Council of Science and Technology, Uttar Pradesh (CSTUP), (Project Id: 4482, CST, U.P. sanction no: CST/D- 7/8, Project Id: 2470, CST, U.P. sanction no: CST/D-1520). B. Bhoi acknowledges support by the Science and Engineering Research Board (SERB) India- SRG/2023/001355. R.K. Nayak acknowledges
computational facilities provided by IIT (BHU) Varanasi.

\bibliographystyle{unsrt}   % citation style
\bibliography{refrence}   % references.bib file

\end{document}